\documentclass[%
aps,
prb,
reprint,
groupedaddress,
amsmath, amssymb,
showpacs
]{revtex4-1}

\usepackage{amsmath}	
\usepackage{graphicx}   
\usepackage{dcolumn}    
\usepackage{bm}         

\newcommand{\cf}{\mathcal{F}}
\newcommand{\tf}{\tilde{\mathcal{F}}}
\newcommand{\vcf}{\bm{\mathcal{F}}}
\newcommand{\vtf}{\tilde{\bm{\mathcal{F}}}}

\begin{document}


\title{Odd-frequency Cooper pairs and zero-energy surface bound states in
superfluid $^3$He}



\author{S. Higashitani}
\author{S. Matsuo}
\author{Y. Nagato}
\author{K. Nagai}


\affiliation{Graduate School of Integrated Arts and Sciences, Hiroshima
University, Kagamiyama 1-7-1, Higashi-Hiroshima 739-8521, Japan}

\author{S. Murakawa}
\author{R. Nomura}
\author{Y. Okuda}

\affiliation{Department of Physics, Tokyo Institute of Technology, Meguro,
Tokyo 152-8551, Japan}


\date{\today}

\begin{abstract}
 We study the odd-frequency Cooper pairs formed near the surface of
 superfluid $^3$He. The odd-frequency pair amplitude is closely related to
 the local density of states in the low energy limit. We derive a formula
 relating explicitly the two quantities. This formula holds for arbitrary
 boundary condition at the surface. We also present some numerical results
 on the surface odd-frequency pair amplitude in superfluid $^3$He-B. Those
 analytical and numerical results allow one to interpret the midgap surface
 density of states, observed recently by transverse acoustic impedance
 measurements on superfluid $^3$He-B, as the manifestation of the surface
 odd-frequency state.
\end{abstract}


\maketitle

\section{Introduction}

The spin and orbital degrees of freedom of the Cooper pairs are customarily
classified into two categories: spin-singlet even-parity and spin-triplet
odd-parity. This classification is based on the requirement that the
equal-time pair amplitude is antisymmetric under simultaneous exchange of
spin and position variables. Berezinskii has proposed a concept of even- and
odd-frequency symmetries associated with the relative-time dependence of the
pair amplitude.\cite{Berezinskii} The even-frequency state is characterized
by the pair amplitude symmetric in frequency, so that it can be subdivided
into the same symmetry classes as in the equal-time case. On the other hand,
the odd-frequency state can have the symmetries of spin-singlet odd-parity
and spin-triplet even-parity, in contrast to the even-frequency state.  The
even-frequency symmetry class includes almost all the paring states so far
discussed. For instance, the spin-singlet $s$-wave state discussed by
Bardeen, Cooper, and Schrieffer for superconducting metals,\cite{BCS} the
spin-singlet $d$-wave state for cuprate superconductors,\cite{HTCS} and the
spin-triplet $p$-wave states for superfluid $^3$He \cite{Helium3} and for
Sr$_2$RuO$_4$ \cite{SrRuO} belong to this category.  The possibility of the
odd-frequency pairing has been proposed in the context of cuprate
superconductors \cite{Balatsky} and heavy Fermion
superconductors.\cite{Fuseya} At present, however, the odd-frequency
superconductor or superfluid has not yet been established experimentally.

Bergeret, Volkov, and Efetov pointed out that the odd-frequency Cooper pairs
are generated by spatial inhomogeneity even without interaction responsible
for odd-frequency pairing.\cite{BVE_PRL} After that, several theoretical
studies have been reported on the odd-frequency states in superconducting
junctions.\cite{TanakaG_PRL,TanakaGKU_PRL,Tanaka_PRB} An intriguing example
of the junctions is a diffusive normal metal/spin-triplet superconductor
system. In this system, the proximity-induced superconductivity in the
diffusive normal metal is dominated by the odd-frequency spin-triplet
$s$-wave Cooper pairs because of isotropization by impurity
scattering.\cite{TanakaG_PRL} Such an anomalous proximity effect is also
expected in superfluid $^3$He partly filled with aerogel.\cite{Seiji_JLTP}
In general, there can coexist even- and odd-frequency Cooper pairs near
interfaces or surfaces owing to broken translational symmetry.  The
inhomogeneous superconducting and superfluid systems therefore provide a
good starting point to develop our understanding of the odd-frequency state.

In this paper, we discuss the odd-frequency Cooper pairs generated near the
surface of superfluid $^3$He. The spin-triplet $p$-wave state of bulk
superfluid $^3$He is arguably the best understood non-$s$-wave
state.\cite{Helium3} It is well established that the non-$s$-wave states
have pronounced sensitivity to surface
scattering.\cite{AdGR,Buchholtz,Buchholtz_RSF,Zhang_PRB,Nagato_pwave,Nagai_JPSJ,
Aoki_PRL,Murakawa_PRL,Murakawa_JPSJ} Unlike the $s$-wave state, the
quasiparticle scattering at the surface causes substantial pair-breaking. As
a result, the gap function varies in space over several coherence lengths
from the surface. In addition, the Andreev bound states are formed at the
surface, yielding a characteristic low energy structure in the surface
density of states below the bulk gap. The surface Andreev bound states have
recently attracted renewed interest from the aspect of their Majorana
nature.\cite{Chung_PRL,Nagato_JPSJ,Murakawa_JPSJ,Volovik_JETP,Tsutsumi} The
purpose of the present work is to demonstrate the existence of the surface
odd-frequency Cooper pairs in superfluid $^3$He.  Recently, Aoki {\it et
al.} \cite{Aoki_PRL} and Murakawa {\it et al.}\
\cite{Murakawa_PRL,Murakawa_JPSJ} performed transverse acoustic impedance
measurements to probe the surface of superfluid $^3$He and found an evidence
of the low energy density of states due to the formation of the surface
Andreev bound states in the B phase of superfluid $^3$He. We show that the
observed midgap density of states can also be interpreted as the
manifestation of the surface odd-frequency Cooper pairs in superfluid
$^3$He-B.

\section{Quasiclassical theory}

Our analysis is based on the quasiclassical Green's function theory of
superfluidity and
superconductivity.\cite{SereneRainer,Nagato_RSM,Eschrig_PRB} We apply it to
a semi-infinite system occupying the space $z > 0$. The surface at $z = 0$
may have atomic-scale irregularities, though we assume it to be
macroscopically flat. The surface roughness of atomic scale, which gives
rise to diffuse quasiparticle scattering, is inevitable in actual systems.

The quasiclassical Green's functions, $g$ and $f$, are a function of $(\hat
p, \epsilon, z)$, where $\hat p$ is a unit vector in the direction of the
Fermi momentum and $\epsilon$ is a complex energy variable. One can express
$g$ and $f$ in the form \cite{Eschrig_PRB}
\begin{align}
 g = i \frac{1 + \cf \tf}{1 - \cf \tf},\quad
 f = \frac{2i}{1 - \cf \tf}\cf,
 \label{gf_def}
\end{align}
where $\cf(\hat p,\epsilon,z)$ and $\tf(\hat p,\epsilon,z)$ are $2 \times 2$
spin-space matrices and the notation $\tilde X$ denotes the transformation
$\tilde X(\hat p, \epsilon, z) = X(-\hat p, -\epsilon^*, z)^*$. The spatial
dependence of $\cf(\hat p,\epsilon,z)$ and $\tf(\hat p,\epsilon,z)$ are
governed by
\begin{align}
 &iv_F\hat p_z \partial_z \cf = -2\epsilon\cf + \Delta(\hat p, z)
 + \cf \Delta(\hat p, z)^\dag \cf,
 \label{diffeq_cf}\\
 &iv_F\hat p_z \partial_z \tf = 2\epsilon\tf - \Delta(\hat p, z)^\dag
 - \tf \Delta(\hat p, z)\tf,
 \label{diffeq_tf}
\end{align}
where $v_F$ is the Fermi velocity and $\Delta(\hat p,z)$ is a gap matrix.
The quasiclassical Green's functions have symmetries
\begin{align}
 &g(\hat p, \epsilon, z) = g(\hat p, \epsilon^*, z)^\dag,
 \label{symrel_g}\\
 &f(\hat p, \epsilon, z) = -f(-\hat p, -\epsilon, z)^T.
 \label{symrel_f}
\end{align}
Here the superscript $T$ denotes matrix transpose.  Equations
\eqref{symrel_g} and \eqref{symrel_f} give the general relation between the
retarded Green's function ($\epsilon = E + i0$) and the advanced one
($\epsilon = E -i0$), and also between the Matsubara Green's functions on
the positive and negative imaginary axes in the complex $\epsilon$ plane.

From $g$, one can calculate the local density of states,
\begin{align}
 n(\hat p,E, z)
 = {\rm Im}\left[\frac{1}{2}{\rm Tr}\,g(\hat p, E + i0, z)\right].
\end{align}
The even- and odd-frequency pair amplitudes are defined in terms of $f$ by
\begin{align}
 \left.
 \begin{array}{l}
  f^{\rm EF}(\hat p, \epsilon, z)\\
  f^{\rm OF}(\hat p, \epsilon, z)\\
 \end{array}
 \right\}
 = \frac{1}{2}[f(\hat p, \epsilon, z) \pm f(\hat p, -\epsilon, z)].
\end{align}

In the following analysis, we consider the unitary states in which
$\Delta\Delta^\dagger \equiv |\Delta|^2$ is proportional to the unit matrix.
The unitary condition is satisfied for all of the spin-singlet states and
for many of spin-triplet states, e.g., those realized in the A and B phases
of superfluid $^3$He \cite{Helium3} and the two-dimensional chiral $p$-wave
state discussed for Sr$_2$RuO$_4$.\cite{SrRuO}

In the unitary states, $\cf$ and $\tf$ for $z \rightarrow \infty$ take the
form
\begin{align}
 \cf^\infty = \frac{\Delta^\infty(\hat p)}{\epsilon +
 i\Omega(\hat p, \epsilon)},\quad
 \tf^\infty = \frac{\Delta^\infty(\hat p)^\dag}{\epsilon +
 i\Omega(\hat p, \epsilon)}.
\end{align}
Here $\Delta^\infty(\hat p) = \Delta(\hat p, z \rightarrow \infty)$ is the
bulk gap matrix and $\Omega(\hat p,\epsilon) = (|\Delta^\infty(\hat p)|^2 -
\epsilon^2)^{1/2}$ in which $|\Delta^\infty(\hat p)|$ corresponds to the
bulk energy gap. The $\cf^\infty$ and $\tf^\infty$ make the right-hand side
of Eqs.\ \eqref{diffeq_cf} and \eqref{diffeq_tf} in the bulk region vanish
and substitution of them into Eq.\ \eqref{gf_def} gives the well-known bulk
solution of the quasiclassical Green's functions, $g^\infty =
\epsilon/\Omega(\hat p, \epsilon)$ and $f^\infty = \Delta^\infty(\hat
p)/\Omega(\hat p,\epsilon)$.

The boundary problem in the semi-infinite system can be solved in the
following way.\cite{Eschrig_PRB} To obtain the incoming (outgoing) solution
$\cf_{\hat p_z < 0}$ ($\cf_{\hat p_z >0}$), the differential equation
\eqref{diffeq_cf} is integrated from right (left) to left (right). Equation
\eqref{diffeq_tf} for $\tf$ is integrated in the opposite direction. The
information on the surface is therefore included in the initial values of
the outgoing solutions ($\cf_{\hat p_z > 0}$ and $\tf_{\hat p_z < 0}$),
while the incoming ones ($\cf_{\hat p_z < 0}$ and $\tf_{\hat p_z > 0}$) are
independent of the boundary condition at the surface apart from the implicit
dependence through $\Delta(\hat p,z)$.

\section{Relation between the pair amplitude and the local density of states}

We now consider the retarded Green's functions ($\epsilon = E + i0$) for the
spin-triplet unitary states and show that the odd-frequency pair amplitude
is closely related to the local density of states in the low energy
limit. The spin-triplet states are characterized by the gap matrix of the
form
\begin{align}
 \Delta(\hat p,z) = \bm{d}(\hat p,z)\cdot\bm{\sigma}i\sigma_2,
\end{align}
where $\bm{\sigma} = (\sigma_1, \sigma_2, \sigma_3)$ is the Pauli matrix.
The $d$-vector in the unitary states can be expressed as the product of a
real vector and a phase factor:
\begin{align}
 \bm{d}(\hat p,z) = \bm{d}_0(\hat p,z) e^{i\varphi_0(\hat p,z)}.
\end{align}

In the absence of the magnetic field, the matrix $\cf$ does not have a
spin-singlet component and we may put
\begin{align}
 \cf = \vcf\cdot\bm{\sigma}i\sigma_2,\quad
 \tf = -i\sigma_2\vtf\cdot\bm{\sigma}.
 \label{cftf_triplet}
\end{align}
It is straightforward to show that the incoming solutions of $\vcf$ and
$\vtf$ for $|E| < |\Delta^\infty(\hat p)|$ can be parametrized as
\begin{align}
 \vcf_{\hat p_z < 0} = -i \hat a e^{i\gamma},
 \quad
 \vtf_{\hat p_z > 0} = i \hat a e^{-i\gamma},
 \label{vf_gam}
\end{align}
where $\hat a$ is a unit vector and $\gamma$ is a real function; those are
determined from
\begin{align}
 &v_F\hat p_z\partial_z\hat a = 2\cos(\gamma - \varphi_{0})
 \left[\bm{d}_{0} - (\bm{d}_{0}\cdot\hat a)\hat a\right],
 \label{diffeq_a}\\
 &v_F\hat p_z\partial_z\gamma = 2\left[E - \sin(\gamma -
 \varphi_{0})(\bm{d}_{0}\cdot\hat a)\right],
 \label{diffeq_gamma}
\end{align}
and the initial values in the bulk region,
\begin{align}
 &\hat a^\infty
 = -{\rm sgn}(\hat p_z) {\bm{d}_{0}^\infty}/{|\bm{d}_{0}^\infty|},\\
 &\gamma^\infty = \varphi_{0}^\infty
 - {\rm sgn}(\hat p_z) \arcsin\left({E}/{|\bm{d}_{0}^\infty|}\right).
 \label{bcon_gamma}
\end{align}
Equations \eqref{gf_def} and \eqref{cftf_triplet} yield $g = g_0 +
\bm{g}\cdot\bm{\sigma}$ and $f = \bm{f}\cdot\bm{\sigma}i\sigma_2$ with
\begin{align}
 &g_0 = \frac{i}{N}\left(1 - \vcf^2\vtf^2\right),\quad
 \bm{g} = -\frac{2}{N}\vcf\times\vtf,
\end{align}
and
\begin{align}
 &\bm{f} = \frac{2i}{N}\left(\vcf - \vcf^2\vtf\right),
\end{align}
where $N = 1 - 2 \vcf\cdot\vtf + \vcf^2\vtf^2$.  From the above expressions
for $g_0$ and $\bm{f}$, one can derive the following relations:
\begin{align}
 g_0\vcf = \frac{1}{2}\left(\bm{f} - \vcf^2\tilde{\bm{f}}\right),
 \quad
 g_0\vtf = -\frac{1}{2}\left(\tilde{\bm{f}} - \vtf^2{\bm{f}}\right).
 \label{g0_vf}
\end{align}
Combining Eqs.\ \eqref{vf_gam} and \eqref{g0_vf}, we obtain
\begin{align}
 (g_0 \hat a)_{(\hat p, E + i0, z)}
 = \frac{i}{2}\left(\bm{f}e^{-i\gamma} + \tilde{\bm{f}}e^{i\gamma}
 \right)_{(\hat p, E + i0, z)}.
 \label{g0_fvec}
\end{align}
Taking the imaginary part of this equation, we get
\begin{align}
 \left.n(\hat p, E, z)\right|_{|E| < |\Delta^\infty(\hat p)|}
 = \left|{\rm Re}\left[\bm{D}(\hat p, E, z)e^{-i\gamma}\right]
 \right|
 \label{ldos_pair}
\end{align}
with
\begin{align}
 \bm{D}(\hat p, E, z)
 &= \frac{1}{2}\left[\bm{f}(\hat p, E + i0, z)
 + \tilde{\bm{f}}(\hat p, E + i0, z)^*\right] \notag \\
 &= \frac{1}{2}\left[\bm{f}(\hat p, E + i0, z)
 - \bm{f}(\hat p, E - i0, z)\right],
\end{align}
where the last equality follows from Eq.\ \eqref{symrel_f}.  At $E = 0$, the
right-hand side of Eq.\ \eqref{ldos_pair} can be written in terms of the
odd-frequency pair amplitude,
\begin{align}
 n(\hat p, 0, z) = \left|
 {\rm Re}\left[\bm{f}^{\rm OF}(\hat p, i0, z)e^{-i\gamma}\right]
 \right|.
 \label{ldos_zero_pair}
\end{align}
It follows that the observation of a finite density of states at zero energy
gives a direct evidence of the existence of the odd-frequency Cooper pairs.

Note that the phase $\gamma$ at $E = 0$ arises essentially due to the
breaking of time-reversal symmetry. This is because the $d$-vector for the
time-reversal invariant states can be taken to be real ($\varphi_0 = 0$) and
then we see from Eqs.\ \eqref{diffeq_gamma} and \eqref{bcon_gamma} that
$\gamma(E = 0) = 0$ at any position $z$. Thus, for the time-reversal
invariant states, Eq.\ \eqref{ldos_zero_pair} is reduced to
\begin{align}
 n(\hat p, 0, z) = \left|
 {\rm Re}\,\bm{f}^{\rm OF}(\hat p, i0, z)
 \right|.
 \label{ldos_zero_pair_TR}
\end{align}
Equation \eqref{ldos_zero_pair_TR} can be applied to superfluid $^3$He-B, as
discussed below.

Before proceeding to discussion on superfluid $^3$He, we briefly mention the
relations similar to the above in the case of spin-singlet states with
$\Delta(\hat p,z) = \Delta_0(\hat p,z) i\sigma_2$ [see also Ref.\
\onlinecite{Barash} in which an analog of Eq.\ \eqref{g0_fvec} for a
spin-singlet superconductor can be found].  For $|E| < |\Delta_0^\infty(\hat
p)|$, we can put $\cf_{\hat p_z < 0} = -i s e^{i\gamma} i\sigma_2$ and
$\tf_{\hat p_z > 0} = i s e^{-i\gamma} i\sigma_2$ with $s = {\rm sgn}[{\rm
Re}\,\Delta_0(\hat p,z)]$. Then, we can arrive, in a similar way to the
spin-triplet case, at
\begin{align}
 n(\hat p, 0, z) =
 \left|{\rm Re}\left[
 f_0^{\rm OF}(\hat p, i0, z)e^{-i\gamma}
 \right]\right|,
 \label{ldos_zero_singlet_pair}
\end{align}
where $f_0^{\rm OF}$ is the odd-frequency spin-singlet pair amplitude, i.e.,
$f^{\rm OF} = f_0^{\rm OF}i\sigma_2$.  For the time-reversal invariant
states, we can again take $\gamma(E = 0)$ to be zero.

Let us turn to the surface odd-frequency state in superfluid $^3$He-B. The
$d$-vector in the semi-infinite superfluid $^3$He-B has the form
\cite{Buchholtz,Buchholtz_RSF,Zhang_PRB,Nagato_pwave,Nagai_JPSJ}
\begin{align}
 \bm{d}
 = (\Delta_\|(z) \hat p_x, \Delta_\|(z) \hat p_y, \Delta_\perp(z) \hat p_z).
\end{align}
Here $\Delta_{\|,\perp}(z)$ are spatially dependent gap functions, both of
which tend for $z \rightarrow \infty$ to the isotropic bulk gap of the B
phase, $|\Delta^\infty(\hat p)| \equiv \Delta_B$. Since
$\Delta_{\|,\perp}(z)$ can be taken to be real, this state is an example of
the time-reversal invariant spin-triplet states.

The $p$-wave gap functions $\Delta_{\|,\perp}(z)$ have the following spatial
structure depending on the boundary condition at the
surface.\cite{Buchholtz,Buchholtz_RSF,Zhang_PRB,Nagato_pwave,Nagai_JPSJ}
When the surface is specular, the perpendicular component $\Delta_\perp(z)$
is strongly suppressed near the surface by the destructive interference
effect due to quasiparticle reflection, while the parallel component
$\Delta_\|(z)$ is slightly enhanced to compensate the condensation energy
lost by the suppression of $\Delta_\perp(z)$. At rough surface causing
diffuse quasiparticle scattering, the parallel component also suffers from
the interference effect, so that both of $\Delta_\|(z)$ and
$\Delta_\perp(z)$ are suppressed. The numerical results for
$\Delta_{\|,\perp}(z)$ determined self-consistently from the gap equation
can be found in Refs.\
\onlinecite{Buchholtz,Buchholtz_RSF,Zhang_PRB,Nagato_pwave,Nagai_JPSJ}.

\begin{figure}[t]
 \includegraphics[width = 8.4cm]{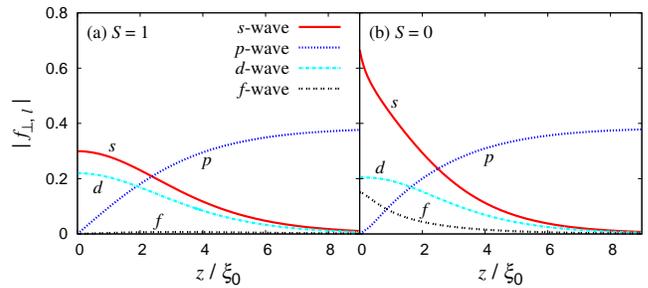}
 \caption{\label{fig1}
 (Color online) Spatial dependence of $|f_{\perp,l}(E+i0,z)|$ at $E =
 0.5\Delta_B$ in the semi-infinite superfluid $^3$He-B at temperature $T =
 0.2T_c$ ($T_c$: superfluid transition temperature). The surface specularity
 $S$ is taken to be (a) $S = 1$ (specular limit) and (b) $S = 0$ (diffusive
 limit). Self-consistently determined gap functions
 $\Delta_{\parallel,\perp}(z)$ are used in the calculations.}
\end{figure}

The self-consistent solution of $\bm{f}$ takes the form
\begin{align}
 \bm{f} = (f_\parallel\cos\phi, f_\parallel\sin\phi, f_\perp),
\end{align}
where $\phi = \arctan(\hat p_y/\hat p_x)$ is the azimuthal angle of $\hat
p$. The pair amplitudes $f_{\|,\perp}$ are a function of $(\hat p_z,
\epsilon, z)$ and are independent of $\phi$ because of the rotational
symmetry around the surface normal.

To illustrate the existence of the surface odd-frequency Cooper pairs in
superfluid $^3$He-B, let us consider the pair amplitude $f_\perp(\hat p_z,
\epsilon, z)$ and its partial-wave components
\begin{align}
 f_{\perp,l}(\epsilon, z) = \langle f_{\perp}(\hat p_z,\epsilon,z)
 P_l(\hat p_z)\rangle_{\hat p},
\end{align}
where $P_l(\hat p_z)$ is the Legendre polynomial and $\langle \cdots
\rangle_{\hat p}$ denotes the angle average over the Fermi surface. It
follows from the general symmetry relation \eqref{symrel_f} that the
even-$l$ (odd-$l$) components have the symmetry of odd (even) frequency. In
Fig.\ \ref{fig1}, we plot the magnitude of the self-consistent solution of
$f_{\perp,l}(E+i0,z)$ at $E = 0.5\Delta_B$ as a function of $z$ scaled by
the coherence length $\xi_0 = v_F/2\pi T_c$. The rough surface effect is
taken into account using the random $S$-matrix theory.
\cite{Nagato_RSM,Nagato_pwave} The surface roughness is parametrized by
specularity $S$ ($0 \leq S \leq 1$; $S=1$ corresponds to the specular
surface and $S \rightarrow 0$ to the diffusive limit where the
quasiparticles are scattered
isotropically).\cite{Murakawa_PRL,Murakawa_JPSJ} The odd-frequency even-$l$
Cooper pairs have substantial amplitudes at the surface in both of the
specular ($S=1$) and diffusive ($S=0$) limits. At specular surface, in
particular, $f_\perp$ has only the odd-frequency components. This is due to
the reflection symmetry $f_\perp(\hat p_z, \epsilon, 0) = f_\perp(-\hat p_z,
\epsilon, 0)$ at the specular surface.

\begin{figure}[t]
 \includegraphics[width = 7.5cm]{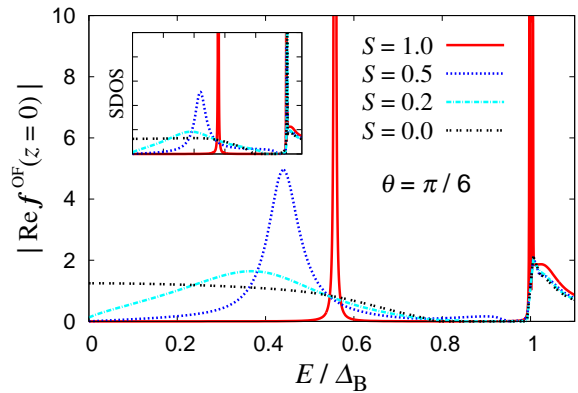}
 \caption{\label{fig2}
 (Color online) $|{\rm Re}\bm{f}^{\rm OF}(\hat p,E+i0,z = 0)|$ in the
 semi-infinite superfluid $^3$He-B at $T = 0.2T_c$ as a function of
 $E/\Delta_B$. The inset is the surface density of states, $n(\hat p, E, z =
 0)$. The two quantities are independent of the azimuthal angle $\phi =
 \arctan(\hat p_y/\hat p_x)$ because of the rotational symmetry around the
 surface normal. The polar angle $\theta = \arccos(\hat p_z)$ is taken to be
 $\pi/6$. The numerical results are shown for several values of the surface
 specularity $S$.}
\end{figure}

\begin{figure}[t]
 \includegraphics[width = 7.5cm]{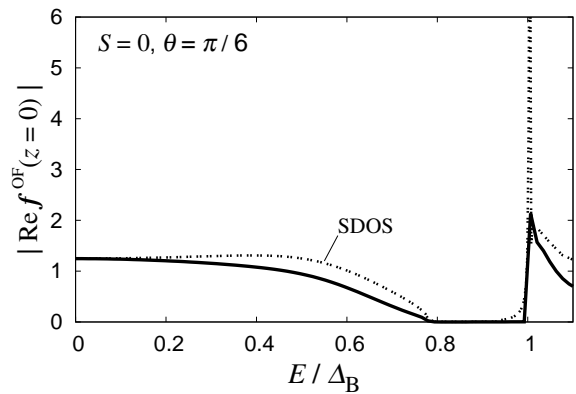}
 \caption{\label{fig3}
 $|{\rm Re}\bm{f}^{\rm OF}(\hat p,E+i0,z = 0)|$ shown in Fig.\ \ref{fig2}
 for the diffusive limit ($S = 0$) is compared with the surface density of
 states, $n(\hat p, E, z = 0)$.}
\end{figure}

In Figs.\ \ref{fig2} and \ref{fig3}, the $E$ dependence of $|{\rm
Re}\,\bm{f}^{\rm OF}|$ for $\epsilon = E + i0$ at the surface of superfluid
$^3$He-B is shown together with the surface density of states (SDOS),
$n(\hat p, E, 0)$.  The two quantities have quite similar $E$ dependence. In
the limit $E \rightarrow 0$, the both coincide exactly with each other, as
we have shown analytically [see Eq.\ \eqref{ldos_zero_pair_TR}].

In Fig.\ \ref{fig2}, the sharp peak (solid line) below the bulk gap
($E/\Delta_B < 1$) corresponds to the Andreev bound state formed near the
specular surface ($S=1$). As the specularity $S$ decreases, the peak is
broadened and the density of states around zero energy increases.  Such low
energy excitations in superfluid $^3$He-B have recently been observed by
transverse acoustic impedance measurements.\cite{Murakawa_PRL,Murakawa_JPSJ}
In the experiment, the specularity is controlled by coating the surface with
thin $^4$He layers. The specularity dependence of the surface midgap density
of states is detected as a distinctive change in a peak structure found in
the frequency dependence of the impedance below the bulk gap
$\Delta_B$.\cite{Nagai_JPSJ,Murakawa_PRL,Murakawa_JPSJ} This experiment
gives a strong evidence of not only the formation of the surface Andreev
bound states but also the existence of the surface odd-frequency Cooper
pairs.

\section{Conclusion}

Based on the experimental results\cite{Murakawa_PRL,Murakawa_JPSJ} of the
transverse acoustic impedance in superfluid $^3$He and on the formula
\eqref{ldos_zero_pair_TR} relating the odd-frequency pair amplitude to the
zero-energy density of states, we have demonstrated that the odd-frequency
Cooper pairs are formed near the surface of superfluid $^3$He-B. The surface
odd-frequency state is the origin of the anomalous proximity effect
predicted for a diffusive normal metal/spin-triplet superconductor junction
\cite{TanakaG_PRL} and an analogous system composed of aerogel and
superfluid $^3$He-B.\cite{Seiji_JLTP} Since the pairing symmetry of bulk
superfluid $^3$He is well established, its surface or interface provides an
ideal environment for studying the physics of the odd-frequency state.

\begin{acknowledgments}
 We would like to thank Y. Tanaka and Y. Asano for helpful discussions.
 This work was supported in part by a Grant-in-Aid for Scientific Research
 (No.\ 21540365) and the ``Topological Quantum Phenomena'' Grant-in-Aid for
 Scientific Research on Innovative Areas (No.\ 22103003) from MEXT of Japan.
\end{acknowledgments}

\providecommand{\noopsort}[1]{}\providecommand{\singleletter}[1]{#1}%

\end{document}